\documentclass[a4paper,11pt,fleqn]{article}
\usepackage[dvips]{graphicx,color}
\usepackage{latexsym}
\usepackage{amssymb}
\usepackage{amsmath}
\usepackage{amsfonts}

\font\helv=cmssbx10

\def\beeta{\pmb{\eta}}

\begin{document}

\title{\bf A pseudo-unitary ensemble of random matrices, PT-symmetry and the Riemann Hypothesis}
\author{ Zafar Ahmed and Sudhir R. Jain\\
{\em Nuclear Physics Division, Van de Graaff Building,}\\
{\em Bhabha Atomic Research Centre, Trombay, Mumbai 400 085, India}}

\date{}

\maketitle

\begin{abstract}
An ensemble of 2 $\times$ 2 pseudo-Hermitian random matrices is constructed that
possesses real eigenvalues with level-spacing distribution exactly as for the
Gaussian unitary ensemble found by Wigner. By a re-interpretation of Connes' spectral
interpretation of the zeros of Riemann zeta function, we propose to enlarge the scope
of search of the Hamiltonian connected with the celebrated Riemann Hypothesis by
suggesting that the Hamiltonian could also be {\helv PT}-symmetric (or pseudo-Hermitian). 
\end{abstract}
\newpage

Riemann Hypothesis (RH) states that all the nontrivial zeros of the Riemann zeta function
have the form $\frac{1}{2}+i\sigma_n$, lying on a line \cite{riemann}. This beautiful
statement got related to mechanics by the conjecture of Hilbert and Polya, as a result,
a search is on for a self-adjoint operator admitting real eigenvalues $\{\sigma _n\}$.
Perhaps the most striking work in this direction is due to Connes \cite{connes} who
constructed a classical eigenvalue problem with a Perron-Frobenius operator and presented
a spectral interpretation of Riemann zeros. In the realm of quantum mechanics, observations
on trace formula have given some important insights and several Hamiltonians have been
discussed \cite{bk1,bk,okubo,others}.

The connection of the RH with random matrix theory (RMT) is very deep. For the statistical
description of level-sequences (or number-sequences) of nuclei, Wigner \cite{mehta}
introduced the subject wherein Hamiltonian matrices were constructed keeping in mind
the underlying symmetries possessed by a physical system. Thus, an even-spin,
time-reversal invariant system belongs to a Gaussian Orthogonal Ensemble (GOE) whereas a
system
violating time-reversal invariance (TRI) belongs to a Gaussian Unitary Ensemble (GUE). That the
sequence $\{\sigma _n\}$ actually has a spectral interpretation first came out of the
seminal work by Montgomery \cite{montgomery} who found the two-point correlation of Riemann
zeros and obtained exactly the result known for GUE. Since then, higher-order correlations
among Riemann zeros have also been shown to correspond to GUE; within GUE, it is known that
two-point correlation guarantees all the higher-order correlations as they are factorisable.
Perhaps the most important single effort after the Montgomery's work is the marathon numerics
by Odlyzko \cite{odlyzko} who has decidedly shown that the distribution has the form
exactly as in GUE.
Due to these works, the Hamiltonians being searched for RH are the ones where time-reversal
invariance is broken \cite{bk,okubo,others}.

Let us focus on two main points on which all the works rest. Firstly, the reality of eigenvalues
of a Hermitian operator and completeness of solutions of the ensuing eigenvalue problem would
guarantee that the RH holds true. Secondly, due to the mathematical and numerical works on
correlations,
it is expected that the Hamiltonian underlying the RH breaks TRI. In this paper, we construct a
pseudo-unitary ensemble of random matrices which has the spacing distribution exactly as in GUE.
Since these systems usually correspond to the physical situation where TRI and parity are not
individually preserved, the finding presented below suggests that the Hamiltonian underlying RH
could also be pseudo-Hermitian. After this demonstration, we shall provide further reasons
that attest to the above statement.

A Hamiltonian {\helv H} is called pseudo-Hermitian \cite{ph,ptph} if
$\beeta${\helv H}$\beeta ^{-1}$ = {\helv H}$^{\dagger}$ for some metric $\eta$.
If $E_m$ and $E_n$ are two eigenvalues of {\helv H}, it is known that \cite{ptph}
\begin{equation}
(E^\ast_m-E_n)~ \langle \Psi^\ast_m |\beeta~\Psi_n \rangle =0,
\end{equation}
implying that if eigenvalues are real and different, the eigenstates are orthogonal
as $\langle \Psi^\ast_m |\beeta \Psi_n \rangle =\in_n \delta_{m,n}$.
If an eigenvalue is complex it will have a zero pseudo-norm as $N_{\beeta}=\langle
\Psi_n^\ast |\beeta \Psi_n\rangle =0.$ The vanishing of pseudo-norm means that the
eigenvector is null. If a Hamiltonian, {\helv H} is symmetric under a joint
action of parity {\helv P} : $x\rightarrow -x$ and time-reversal {\helv T} : $i \rightarrow -i$,
i.e, ({\helv PT}) {\helv H}  ({\helv PT})$^{-1}$={\helv H} \cite{pt} then we have real eigenvalues if the eigenstates, $\Psi_n$
are also the eigenstates of {\helv PT}, otherwise the eigenvalues are complex conjugate pairs.
For {\helv PT}-symmetric Hamiltonians we have \cite{ahmed}
\begin{equation}
(E^\ast_m-E_n)~ \langle \Psi^{\mbox{\helv PT}}_m |\Psi_n \rangle=0.
\end{equation}
When eigenvalues are real and distinct, the eigenstates  are orthogonal
as $\langle \Psi^{\mbox{\helv PT}}_m |\Psi_n\rangle = \in_n \delta_{m,n}$.

Remarkably, {\helv PT}-symmetric Hamiltonians are found to be pseudo-Hermitian :
{\helv P H P}$^{-1}$ = {\helv H}$^\dagger$. Pseudo-Hermiticity has been recast 
\cite{ptph} in terms of {\helv PT}-symmetry.
Given a pseudo-Hermitian Hamiltonian, one can construct generalized {\helv P} and 
{\helv T}.

The operator {\helv D}=
$e^{i\mbox{\helv H}}$ is pseudo-unitary in accordance with \cite{aj}
\begin{equation}
\mbox{\helv D}^{\dagger}
= \beeta \mbox{\helv D}^{-1}\beeta ^{-1}.
\end{equation}
The eigenvalues of {\helv D} are either on the unit circle or of the type :
$|\lambda_1 \lambda_2|=1.$
It is only recently that a random matrix theory has been presented for a statistical
study of pseudo-Hermitian Hamiltonians \cite{aj}. Only 2$\times$2 matrices
have been studied. To summarize briefly, two cases are found : one with a linear (with
more slope than that of goe) level repulsion and the other where as the spacing
$s$ becomes small, level-spacing distribution $\sim s\log \frac{1}{s}$ \cite{aj}.
We call these ensembles \cite{cz} as Gaussian pseudo-orthogonal ensemble (GPOE) and Gaussian
pseudo-unitary ensemble (GPUE), respectively. The essence of these two results is
that they show much weaker level-repulsion at small spacings than those of the known
ensembles of Wigner and Dyson.

Let us now consider the Hamiltonian matrix
\begin{eqnarray}
\mbox{\helv H} = \{\mbox{\helv H}_{ij}\}&=
\left[\begin{array}{cc}a+b&(c+id)/\epsilon \\(c-id)\epsilon&a-b\end{array}\right],
\end{eqnarray}
$a, b, c, d$ being real. This is pseudo-Hermitian with respect to a metric
\begin{eqnarray}
\beeta = \left[\begin{array}{cc}\epsilon&0\\0&1/\epsilon\end{array}\right]
\end{eqnarray}
which gives rise to a positive definite pseudo-norm (1)
It is due to this property such Hamiltonians as (4) are called quasi-Hermitian
(see Scholtz et al. in \cite{ph}).
The eigenvalues of {\helv H} are given by
\begin{equation}
E_{\pm} = a \pm \sqrt{b^2+c^2+d^2}.
\end{equation}
Consider that the matrix {\helv H} is drawn from an ensemble
of random matrices with a Gaussian distribution given by \cite{mehta}
\begin{equation}
P(\mbox{\helv H}) = {\cal N} e^{- \frac{1}{2\sigma ^2}~tr~\mbox{\helv H}^{\dagger}
\mbox{\helv H}}.
\end{equation}
Accordingly, the joint probability distribution of $a, b, c, d$ is
\begin{equation}
P(a, b, c, d) \sim \exp \left[-\frac{1}{\sigma ^2}\left(a^2+b^2+
\left(\epsilon ^2+\epsilon ^{-2}\right)\frac{(c^2+d^2)}{2}\right)\right].
\end{equation}
We know that three-parameter unitary matrix, {\helv U}$(\theta,\phi,\psi)$
\begin{eqnarray}
\mbox{\helv U} = \left[\begin{array}{cc} e^{i\psi}\cos \theta &-\sin \theta e^{i\phi}
\\ \sin \theta e^{-i\phi} &e^{-i\psi}\cos \theta \end{array}\right].
\end{eqnarray}
constitutes a Lie group. More importantly, the unitary matrix {\helv U} can
generate all the Hermitian $2 \times 2$ matrices of the general type ($\epsilon=1,$
in (5)) with any arbitrary value (including zero) of $\psi$. Only two continuous
parameters $(\theta,\phi)$ suffice for this purpose. Inspired by this, we construct the
following matrix {\helv D}:
\begin{eqnarray}
\mbox{\helv D} = \left[\begin{array}{cc}\cos \theta &-\sin \theta e^{i\phi}/\epsilon \\
\sin \theta e^{-i\phi}\epsilon &\cos \theta \end{array}\right], 
\end{eqnarray}
which is pseudo-unitary with respect to $\beeta$.
As per our design {\helv D} matrix would generate all possible {\helv H} of the type (4) as
\begin{equation}
\mbox{\helv D} \mbox{~diag}(E_+,E_-)\mbox{\helv D}^{-1} = \mbox{\helv H},
\end{equation}
which gives us the following relations :
\begin{eqnarray}
&~&a = \frac{E_++E_-}{2}, ~~ b = \frac{E_+-E_-}{2}\cos 2\theta , \nonumber \\
&~&c = \frac{E_+-E_-}{2}\sin 2\theta \cos \phi, ~~ d = \frac{E_+-E_-}{2}\sin 2\theta \sin \phi .
\end{eqnarray}
Writing $\epsilon = e^{-\gamma}$, and calling $t=E_++E_-$, $s=E_+-E_-$, we have
$P(a,b,c)$ going over to
\begin{equation}
P_{\gamma}(s, t, \theta , \phi ) \sim \exp
\left[-\frac{t^2}{4\sigma ^2}-\frac{s^2}{4\sigma ^2}\cos^2 2\theta-\frac{s^2}
{\sigma ^2}\cosh 2\gamma \cos ^2\theta \sin ^2\theta \right]
\end{equation}
via a Jacobian, ${\cal J}=\frac{s^2 \sin 2\theta}{4}$. Next, integrating over $t,
\theta $ and $\phi$, we have the un-normalised nearest-neighbour spacing distribution given by
\begin{equation}
P_{\gamma}(s) \sim s\exp \left(-\frac{p^2s^2}{\sigma ^2} \right)
\mbox{Erfi}\left(\frac{q}{2\sigma}s \right)
\end{equation}
where
\begin{equation}
\mbox{Erfi}(x)=\frac{x}{\sqrt{\pi}}\int_{-1}^{+1} dy e^{x^2 y^2},
p=\sqrt{\cosh 2\gamma}/2, q=\sqrt{\cosh 2\gamma-1}/2
\end{equation}
We write now the normalized nearest-neighbour spacing distribution in terms of
a dimensionless variable, $x=\frac{s}{\langle s \rangle} $ where $\langle s \rangle $ is
the mean level spacing. With 
\begin{eqnarray}
\alpha_{\gamma}=\frac{2}{\sqrt{\pi}}\left( 1 + \frac{\tanh ^{-1} (q/p) }{4pq} \right), \nonumber \\ 
P_{\gamma}(x) = \frac{\alpha _{\gamma }^2\cosh 2\gamma}{4 } ~x~ e^{-p^2\alpha^2_{\gamma}x^2}
\left[ \frac{\mbox{Erfi}(\alpha_{\gamma} q x)}{q} \right].
\end{eqnarray}
Note that the limiting value of the square-bracketted term is  
$\frac{2x\alpha_0}{\sqrt{\pi}}$ and $\alpha_{0}=\frac{4}{\sqrt{\pi}}$. 
For an arbitrarily small $\gamma$, the matrix {\helv H} is pseudo-Hermitian. 
Actually, even for $\gamma$
as much as 1/2, the difference between $P_{\gamma}(x)$ and $P_{\mbox{\small GUE}}(x)=
\frac{32 }{ \pi^2} x^2 e^\frac{-4x^2 }{ \pi}$ is hardly appreciable (see Fig. 1(a)).

\begin{figure}
\includegraphics[width=13.5cm,height=6cm]{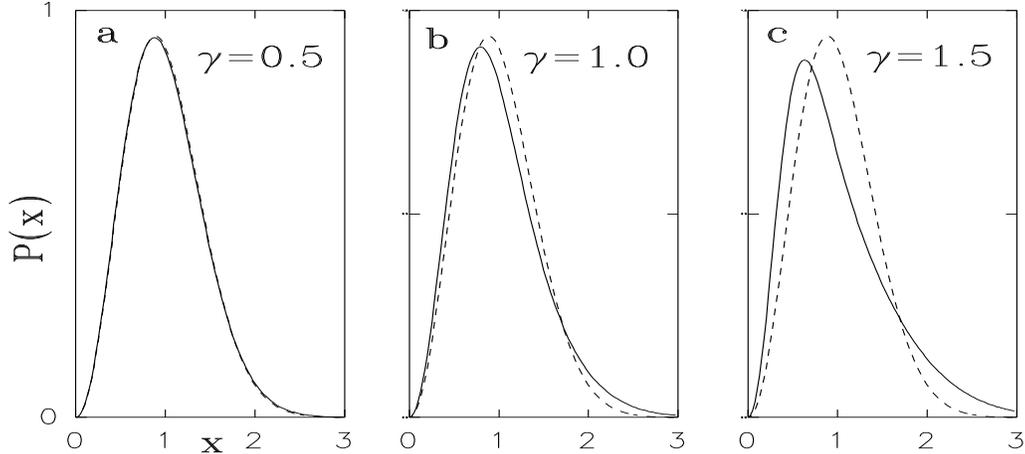}
\caption {Plot of $P_{\gamma}(x)$ (solid line) (16) for three values 
of $\gamma$. Dashed line denotes $P_{GUE}(x)$.}
\end{figure}

Returning to the discussion of the RH, with this example-ensemble, the scope of
search of the Hamiltonian for the RH widens. The Hamiltonian in question relevant
for the RH could be pseudo-Hermitian. Although the example presented is non-generic,
so could the Riemann Hamiltonian be. There is nothing that suggests generic nature of
the Hamiltonian, particularly in the light of our illustrative example.

Our suggestion is also well supported by the spectral interpretation
of the Riemann zeros ensuing from Connes' work \cite{connes}. According to
Connes, the zeros form an absorption spectrum in the sense that the wavefunctions
corresponding to the eigenvalues $\sigma _n$ is ``zero''. We know that the eigenvalues
of a pseudo-Hermitian operator are either real or complex-conjugate pairs.
Thus, {\it we suggest the possibility of the Riemann zeros
$\{\frac{1}{2}\pm i\sigma _n\}$ to be the complex-conjugate-pair eigenvalues of
an unknown pseudo-Hermitian operator where it would be automatically guaranteed
that the eigenvectors are null.} Our paper suggests this central message.

Let us demonstrate our point heuristically by taking a simplistic and
trivially {\helv PT}-symmetric Hamiltonian as
\begin{equation}
\mbox{\helv H}_{\mbox{\helv PT}}=-i\mbox{\helv x p}.
\end{equation}
The eigenvalues like $\frac{1}{2} \pm i t_n$ will be supported with $\Psi_n(x)=
N x^{-\frac{1}{2}\mp i t_n}$ such that $\Psi_n(\pm \infty)=0$ and 
{\helv H}$_{\mbox{\helv PT}}\Psi_n=(\frac{1}{2}\pm it_n) \Psi_n(x)$.  
Very importantly notice that the eigenvalues are complex conjugate
pairs and the eigenfunctions of {\helv H}$_{\mbox{\helv PT}}$ are not the simultaneous eigenstates of
the antilinear operator, {\helv PT}. As stated earlier, this situation is referred to
as spontaneous breaking of {\helv PT}-symmetry \cite{pt}. 
Check that {\helv PT}$\Psi_n(x)=\Psi^\ast_n(-x) \ne c \Psi_n(x)$.
Next, whether $\frac{1}{2}+it_n$ are bona fide discrete eigenvalues and whether $t_n$
would coincide with $\sigma_n$ (RZs) are of course the most crucial questions.

Our simple Hamiltonian in (17) mimics the Hamiltonian of Berry and Keating
\cite{bk1}
\begin{equation}
\mbox{\helv H}_{\mbox{\small BK}} = \mbox{\helv xp}-\frac{i}{2}
\end{equation}
which, in turn, has been inspired by the work of Connes.
This is Hermitian and it also breaks time-reversal symmetry.
Berry and Keating \cite{bk1,bk} have studied the the semiclassical trace formula for their
Hamiltonain (18) {\it vis-a-vis}
very interesting properties of $\zeta(z)$ and reported a shortcoming of (18) in this regard.
Also they speculated \cite{bk1} that the Hamiltonian (18) along with and extraordinary boundary
condition on the wavefunction would yield  $\pm \sigma_n$ as eigenvalues.
This boundary condition is unfortunately not known so far. Also, if $\sigma_n$
is an eigenvalue, apparantly there is nothing to ensure that $-\sigma_n$
would also be an eigenvalue.

When we diagonalize the  matrix for {\helv H}$_{\mbox{\small BK}}$ using  the one-dimensional Harmonic
Oscillator basis by using the creation and innhallation operators as : {\helv x}=({\helv a}+{\helv a}$^\dagger)/
\sqrt{2}$ and {\helv p}=i({\helv a}$^\dagger $-{\helv a})/ $\sqrt{2}$, we find that eigenvalues very
crucially depend upon the size of the basis (say $N$)! This is how we conclude
that {\helv H}$_{\mbox{\small BK}}$ does not even possess a discrete spectrum. So is the fate of our
toy model {\helv H}$_{\mbox{\helv PT}}$ (17), this however is only heuristic. These simple findings are
for the most ordinary boundary condition where the eigenfunctions vanish at $\pm \infty$.
The real part turns out to be $1/2$, and this would change as soon as the boundary conditions are
disturbed.

The classical analogue of the Hamiltonian (18) is known to be  scaling type 
(as x$\rightarrow $ K x, p$\rightarrow $ p/K), therefore the complex scaling of 
co-ordinate can not be employed to study its resonances.
However, its canonically-transformed Hamiltonian, $H=(p^2-x^2)/2$ is very well-
studied \cite{bv} for its resonances and these are well-known as $\pm i(n+\frac{1}{2})$ (with
$\hbar=1$) not showing any connection with $\sigma_n$.

Okubo \cite{okubo} has considered the Hamiltonian,
\begin{equation}
\mbox{\helv H}_{\mbox{\small Okubo}} = - \mbox{\helv p}_x\mbox{\helv p}_y - 
(1-\beta )\mbox{\helv xp}_x - \beta \mbox{\helv yp}_y + \frac{i}{2}
\end{equation}
that is both Hermitian and time-reversal breaking. It is in two-dimensional Euclidean
space with boundary conditions on the eigenfunctions :\begin{equation}
\mbox{\helv H}_{\mbox{\small Okubo}}\psi (x,y) = \lambda \psi (x,y); ~~\psi (x,0) = 0,
\end{equation}
and $\psi (x,y)$ rapidly decreasing at infinity. We have again constructed the
Hamiltonian matrix for (19) in the harmonic oscillator basis and diagonalised
the matrices to find the eigenvalues. In $x$ it is a H.O. and in $y$ it is 
half-H.O. model as per \cite{okubo}. 
We find that the eigenvalues are not stable with the size of the matrices,
thus indicating that there is no discrete spectrum supported by the Hamiltonian.
However, the possibility of the Riemann zeros to correspond to resonances remains with
this Hamiltonian. It may again be non-trivial to investigate its resonances.

The Hamiltonian suggested by Castro et al. \cite{Castro} is like 
{\helv H}$_{\mbox{\small CGM}}=i${\helv x p + g(x)},
where {\helv g(x)} is a special function and $x>0$. Once again, by finding the matrix elements
employing half H.O. basis, we do not find discrete specrum as the eigenvalues keep changing
with the size, $N$, of the basis.

We have found that the popular Hamiltonians in the context of the
RH do not even possess a discrete spectrum. From the random matrix ensemble of 
pseudo-Hermitian matrices
presented here exhibiting GUE statistics and by a re-interpretation of Connes' work,
we have suggested that the Hamiltonian relevant to the RH could be pseudo-Hermitian.

\end{document}